\documentclass[%
 reprint,
superscriptaddress,
showpacs,
nofootinbib,
 amsmath,amssymb,
 aps,
pra,
]{revtex4-1}

\usepackage{graphicx}% Include figure files
\usepackage{dsfont}

\usepackage{color}
\usepackage[english]{babel}
\usepackage[T1]{fontenc}
\begin{document}

\title{Two-body correlations and natural orbital tomography in ultracold 
      bosonic systems of definite parity}

\author{Sven Kr\"onke}
	\email{skroenke@physnet.uni-hamburg.de}
	\affiliation{Zentrum f\"ur Optische Quantentechnologien, Universit\"at
Hamburg, Luruper Chaussee 149, 22761 Hamburg, Germany}
\author{Peter Schmelcher}
	\email{pschmelc@physnet.uni-hamburg.de}
	\affiliation{Zentrum f\"ur Optische Quantentechnologien, Universit\"at 
Hamburg, Luruper Chaussee 149, 22761 Hamburg, Germany}
	\affiliation{The Hamburg Centre for Ultrafast Imaging, Universit\"at 
Hamburg, Luruper Chaussee 149, 22761 Hamburg, Germany}

\date{\today}

\begin{abstract}
The relationship between natural orbitals, one-body coherences and
two-body correlations is explored for bosonic many-body systems of 
definite parity with two occupied single-particle states. We show that the
strength of local two-body correlations at the parity-symmetry center
characterizes the number state distribution and controls the
structure of non-local two-body correlations.
A recipe for the experimental reconstruction of the natural orbital
densities and quantum depletion is derived. These insights into 
the structure of the many-body
wave-function are applied to the predicted quantum-fluctuations induced decay of
dark solitons.
\end{abstract}

\maketitle
%%%%%%%%%%%%%%%%%%%%%%%%%%%%%%%%%%%%%%%%%%%%%%%%%%%%%%%%%%%%%%%%%%%%%%%%%%%%%%
\section{Introduction}
In an ideal Bose-Einstein condensate, 
all bosons occupy the same
single-particle state $\phi_0({\bf r})$, whose
density can directly be inferred
from an absorption image measurement of the reduced one-body density
$\rho_1({\bf r})$ 
\cite{Bose-Einstein_Condensation_Dilute_Gases_Pethick_Smith_2008}. Yet in a
non-ideal world, not only do interactions between the atoms affect the
shape of the condensate wave-function $\phi_0({\bf r})$ but also
bring more single-particle
orbitals into play, even at zero temperature, such that $\rho_1({\bf r})$ equals
an incoherent superposition of their densities in general. Theoretically, 
the many-body state can be
characterized by the natural orbitals (NOs) $\phi_i({\bf
r})$ \cite{loewdin_norb55}, i.e.\ eigenvectors of the reduced one-body density
operator $\hat\rho_1$, and their populations, i.e.\ the corresponding
eigenvalues: 
Given a sufficiently large weight, the NO of largest population is
identified with the condensate wave-function and quantum depletion
manifests itself in the population of other NOs 
\cite{Onsager_Penrose_BEC_liquid_He_PR_1956}.
As a matter of fact, correlation effects can be traced
back to both the occupation number distribution of the NOs and their spatial
shape, allowing for a microscopic understanding of various phenomena.
Examples for such phenomena are the Mott-insulating
phase, where e.g.\ an intra-well Tonks-Girardeau transition for a filling factor
of two can be understood in the NO framework
\cite{zoo_of_quantum_phases_and_excitations_in_opt_lattices_Alon_PRL2005,
corr_vs_commensurability_effects_finite_bos_sytems_1d_lattices_Brougos_Zoellner}, 
the quantum-fluctuations induced decay of
dark solitons \cite{quantum_depletion_of_dark_soliton_anomalous_mode_PRA2002,
quantum_entangled_dark_solitons_in_optical_lattices_Carr_PRL09,
delande_many-body_2014,kronke_many-body_2014}, where a NO of particular shape is
dominantly responsible for the soliton contrast reduction in the reduced
one-body density, and
fragmented condensates
\cite{fragmentation_of_BEC_in_multiwell_3d_traps_Alon_PhysLettA_2005,
red_density_matrices_and_coherence_of_trapped_bosons_Sakmann_PRA2008,
soliton_fragmenton,
swift_loss_coherence_Cederbaum_PRL2011}, which are defined as many-body states
with two or more macroscopically occupied NOs.
While there are proposals for the detection of fragmentation and its 
degree \cite{kang_revealing_2014,streltsova_interferometry_2014}, 
the one-body density $\rho_1({\bf
r})$ has, to the best of our knowledge, not yet been unraveled into 
the contributions $|\phi_i({\bf r})|^2$
of the individual NOs by means of a measurement protocol.

In principle, the NOs can be obtained from a tomographic
reconstruction of the reduced one-body density matrix $\rho_1({\bf r,\bf r'})$ and
diagonalization.
Such quantum state tomography is a well established technique
for qubit systems and quantized light fields 
\cite{quantum_state_estimation_rev2004,lvovsky_continuous-variable_2009}.
For interacting ensembles of ultracold atoms, 
information
about the off-diagonal elements $\rho_1({\bf r,\bf r'})$ can
be extracted experimentally from the contrast of interfering slices
coupled out 
a trapped Bose gas 
\cite{bloch_measurement_2000,ritter_observing_2007,donner_critical_2007} or 
homodyning in uniform systems, where one generates copies of one's
system by Bragg pulses and let them interfere
\cite{hagley_measurement_1999,navon_critical_2015}. Furthermore, there are
theoretical proposals for the $\rho_1({\bf r,\bf r'})$
reconstruction based on 
Raman pulse sequences \cite{duan_detecting_2006}, variable
time-of-flight (ToF) measurements \cite{zhang_tomography_2009}, heterodyning
with an auxiliary Bose-Einstein condensate \cite{niu_imaging_2006} or Jaynes
principle of maximum entropy \cite{mouritzen_tomographic_2005}.
Yet due to its non-local
character, it is notoriously difficult to infer $\rho_1({\bf r,\bf r'})$
experimentally, in particular for non-uniform systems.

Rather than aiming at a completely general reconstruction scheme for the NOs,
we focus here on bosonic many-body systems of definite parity
with two occupied single-particle states. Assuming only two occupied
orbitals constitutes the simplest, natural extension for bosons 
beyond the mean-field approximation and is physically justified in various
situations.
For this class of systems,
we derive an experimentally accessible reconstruction recipe, in which
density-fluctuation measurements play a decisive role, and also 
gain insights into generic properties of two-body correlations. 
In particular, we show how the character of
the number state distribution function, the NO densities, 
the structure of two-body correlations and the relationship between
one-body coherences and non-local two-body correlations
crucially depend on the strength of two-body correlations at the
parity-symmetry center. All these relations are derived exactly from the structure
of the many-body wave-function. In order to show the importance
of our results as well as their validity when further NOs are slightly
populated, we apply our 
analytical methodology to the analysis of numerical {\it ab-initio}
data of the quantum-fluctuations induced decay of dark solitons
obtained by the Multi-Configuration
Time-Dependent Hartree 
Method for Bosons (MCTDHB) \mbox{
\cite{MCTDHB_Cederbaum,kronke_non-equilibrium_2013,cao_multi-layer_2013}}
and address also the impact of finite experimental resolution.

This work is organized as follows: First, the setup is introduced in Sec.\
\ref{sec:setup}. Thereafter, we discuss the NO decomposition of the reduced
two-body density in Sec.\ \ref{sec:rho2_decomp}, which forms the basis for the
reconstruction of the odd and even parity NO density
as well as for the characterization of the spatial
structure of two-body correlations in Sections
\ref{sec:rectr_odd}, \ref{sec:rectr_even} and  \ref{sec:struct_corr},
respectively. Finally, we
apply our insights to decaying dark solitons in Sec.\ \ref{sec:appl} and
conclude with
Sec.\ \ref{sec:concl}.
%%%%%%%%%%%%%%%%%%%%%%%%%%%%%%%%%%%%%%%%%%%%%%%%%%%%%%%%%%%%%%%%%%%%%%%%%%%%%%%%
\section{Setup}\label{sec:setup}
% {\em Setup.---} 
In the following, we consider a system of $N$ bosons which are energetically or
dynamically restricted to occupy only two single-particle states of opposite
parity, $\hat \pi|\phi_i\rangle=(-1)^i|\phi_i\rangle$, $i=0,1$. 
Here, $\hat\pi$ denotes the single-particle parity operator which inverts
either all coordinates or only $x$. For simplicity, we suppress
the $y$, $z$ arguments in the position representation and remark that our
results are valid for one-, two- and three-dimensional quantum gases.
The $N$-body state is assumed to possess a
definite $N$-body parity, $\bigotimes_{r=1}^N\hat\pi_r|\Psi\rangle =
\Pi|\Psi\rangle$, $\Pi\in\{-1,1\}$ with $\hat\pi_r$ acting on the $r$-th atom, 
as it is the case for a non-degenerate
ground state of a parity-symmetric many-body Hamiltonian e.g.\ of a bosonic
Josephson junction \cite{a_bosonic_josephson_junction_Oberthaler_JPB2007}. 
For this class of systems, the many-body state is of the form
\begin{equation}\label{eq:wfn}
 |\Psi\rangle=\sum_{k=0}^K A_k|n_0(k),N-n_0(k)\rangle,
\end{equation}
where $|n_0,n_1\rangle$ denotes a number state with $n_i$ bosons in
$|\phi_i\rangle$. In the cases of $N$ even and $\Pi=1$ or $N$ odd and $\Pi=-1$, 
the correct parity is ensured by ${n_0(k)=2k}$. Otherwise, 
$n_0(k)=2k+1$ has to be chosen. In all cases, $K$ denotes the largest 
integer with $n_0(K)\leq N$. By tracing out $N-1$
bosons, one obtains for the reduced one-body density operator
$\hat\rho_1=\Delta|\phi_0\rangle\!\langle\phi_0|+
(1-\Delta)|\phi_1\rangle\!\langle\phi_1|$ so that the NOs are given by
$|\phi_i\rangle$. Here, we have introduced the average fraction of bosons in
the even orbital, $\Delta=\overline{n}_0/N$, where $\overline{(...)}$ 
denotes the average w.r.t the number state probability distribution $|A_k|^2$.
Thus, the quantum depletion equals $\min\{\Delta,1-\Delta\}$ and
the reduced one-body density is given by the incoherent superposition
\begin{equation}\label{eq:rho1}
\rho_1(x)\equiv\langle x|\hat\rho_1|x\rangle=\Delta|\phi_0(x)|^2+
(1-\Delta)|\phi_1(x)|^2. 
\end{equation}

By measuring $\rho_1(x)$ and the real-valued off-diagonal elements
$\rho_1(x,-x)\equiv\langle x|\hat\rho_1|-x\rangle$ for all $x$, one
could in principle reconstruct the NO densities and $\Delta$
without knowledge about the full density matrix $\rho_1(x,x')$. As a consequence
of the NO parities, one finds
$|\phi_{0/1}(x)|^2\propto \rho_1(x)\pm\rho_1(x,-x)$ and $\Delta=[1+\int
{\rm d}x\,\rho_1(x,-x)]/2=[1+{\rm tr}(\hat\pi\hat\rho_1)]/2$, which links
$\Delta$ to the average single-particle parity.
The drawback of this scheme, however,
lies in the fact that it requires precise knowledge about $\rho_1(x,-x)$
for all $x$, which is a challenging quantity to measure.
%%%%%%%%%%%%%%%%%%%%%%%%%%%%%%%%%%%%%%%%%%%%%%%%%%%%%%%%%%%%%%%%%%%%%%%%%%%%%%%%
\section{Results}
\subsection{NO decomposition of the two-body density}\label{sec:rho2_decomp}
% {\em NO decomposition of the two-body density.---} 
Since two-body correlations will indeed provide us an alternative pathway
to the NO reconstruction, we investigate here how the structure of the many-body
wave-function (\ref{eq:wfn}) manifests itself in absorption image noise
correlations. The latter have theoretically been proven to give valuable
physical insights in particular for low-dimensional systems
\cite{altman_probing_2004,sykes_spatial_2008,mathey_noise_2008,
mathey_noise_2009,imambekov_density_2009,singh_noise_2014} and are measurable both after ToF and
{\it in situ} nowadays
\cite{folling_spatial_2005,schellekens_hanbury_2005,manz_two-point_2010,
hung_extracting_2011,endres_observation_2011,cheneau_light-cone-like_2012,
endres_single-site-_2013,jacqmin_sub-poissonian_2011,
armijo_mapping_2011,armijo_direct_2012,perrin_hanbury_2012,noise_corr_exp_foelling2014}. 
For this purpose, we derive the two-body density
$\rho_2(x_1,x_2)\equiv\langle\hat\psi^\dagger(x_1) \hat\psi^\dagger(x_2)
\hat\psi(x_2) \hat\psi(x_1)\rangle/[N(N-1)]$, where $\hat\psi(x)$ denotes
the bosonic field operator: 
\begin{eqnarray}\label{eq:rho2}	
&&\rho_2(x_1,x_2)=2\Re\left(\alpha\,
\phi_{11}(x_1,x_2)\phi_{00}^*(x_1,x_2)\right)+\\\nonumber
&&+\beta|\phi_{00}(x_1,x_2)|^2
+2\gamma|\phi_{01}(x_1,x_2)|^2
+\delta |\phi_{11}(x_1,x_2)|^2,
\end{eqnarray}
with $\phi_{ij}(x_1,x_2)$ abbreviating the normalized symmetrization of the
Hartree product $\phi_i(x_1)\phi_j(x_2)$. In the following derivation, we will
eliminate the off-diagonal term with the coefficient $\alpha$, which is a function
of the coherences $A^*_{k+1}A_k$ between the respective number states, by virtue
of the parity symmetry. The second coefficient is related to the second
moment of the number state distribution $|A_k|^2$ via
$\beta=[\overline{n^2_0}-\overline{n}_0]/[N(N-1)]$ and determines the remaining
coefficients $\gamma=\Delta-\beta$ and $\delta=1+\beta-2\Delta$.

Assuming a finite central density, $\rho_1(0)>0$, we
calculate the two-body correlation function
$g_2(x_1,x_2)\equiv\rho_2(x_1,x_2)/[\rho_1(x_1)\rho_1(x_2)]$
\cite{glauber_quantum_1963,naraschewski_spatial_1999} at the symmetry center
\begin{equation}\label{eq:g2_00}
 g_2(0,0) = \frac{\beta}{\Delta^2}=\frac{N}{N-1}\left(1+\frac{{\rm
var}(n_0)-\overline{n}_0}{\overline{n}_0^2} \right),
\end{equation}
where ${\rm var}(n_0)=\overline{n^2_0}-\overline{n}_0^2$. By measuring the
central density and its fluctuations, one can in principle deduce
$g_2(0,0)$ and, thereby,
characterize the number state distribution in the categories Poissonian, sub-
and super-Poissonian. 
Similarly, a measurement of the $n$-th order
correlation function $g_n(x_1=0,...,x_n=0)$ gives insights into the $n$-th
moment of the number state distribution (cf.\ e.g.\ the
experiment \cite{armijo_probing_2010}
for $n=3$). As we will show, the strength of two-body correlations at
the symmetry center constitutes a key parameter, which controls both the
reconstruction of the NO densities and the relationship between local and
non-local two-body correlations. Regarding the impact of an unavoidably
given finite 
experimental resolution, we refer to the discussion at the end of Sec.\
\ref{sec:appl}.
%%%%%%%%%%%%%%%%%%%%%%%%%%%%%%%%%%%%%%%%%%%%%%%%%%%%%%%%%%%%%%%%%%%%%%%%%%%%%%%%
\subsection{Reconstruction of odd NO}\label{sec:rectr_odd}
% {\em Reconstruction of odd NO.---} 
In order to eliminate the term proportional to $\alpha$ in (\ref{eq:rho2}), 
we make use of
$\phi_1(0)=0$ and consider the non-local 
two-body correlations
$g_2(x,0)=[\beta |\phi_0(x)|^2+\gamma|\phi_1(x)|^2]/[\Delta\rho_1(x)]$. After
substituting the density of the even NO $|\phi_0(x)|^2$ via (\ref{eq:rho1})
and employing (\ref{eq:g2_00}), we obtain
\begin{equation}\label{eq:phi1_reconstr}
 |\phi_1(x)|^2=\frac{g_2(0,0)-g_2(x,0)}{g_2(0,0)-1}\rho_1(x),
\end{equation}
which holds\footnote{In the limit $g_2(0,0)\rightarrow1$, the right hand
side of (\ref{eq:phi1_reconstr}) tends to $|\phi_1(x)|^2$ such that no
information about the NO is gained. This fact is also reflected by $g_2(0,0)=1$
implying $g_2(x,0)=1$ (cf.\ Sec.\ \ref{sec:struct_corr}).} for non-trivial
two-body correlations at the symmetry
center, i.e.\ $g_2(0,0)\neq 1$. Under this condition, we have thus shown that
the density of the odd NO is proportional to the total reduced one-body 
density spatially modulated by the strength of non-local two-body
correlations between the symmetry center and the position $x$ of interest. 
This relationship constitutes a key result of this work since it provides an
explicit reconstruction scheme for the microscopic quantity $|\phi_1(x)|^2$
in terms of the measurable densities $\rho_1(x)$ and $\rho_2(x,0)$. 
In particular, this simple reconstruction recipe does not require to measure 
the off-diagonal elements $\rho_1(x,-x)$.
For the ground state
of a parity symmetric Hamiltonian with 
short-range interactions, the 
many-body 
parity turns out to be even \cite{feynman_lectures_stat72} and, for not too strong
interactions, most of the bosons occupy
the NO of even parity. Thus, (\ref{eq:phi1_reconstr}) gives experimental
access to the density of the orbital whose population is
responsible for quantum
depletion.
Besides, the positive semi-definiteness of $|\phi_1(x)|^2$ and
$\rho_1(x)$ implies that
(anti-)bunching at the symmetry center yields $g_2(0,0)$ as an upper (lower) bound
for the non-local correlation $g_2(x,0)$.
\begin{figure*}[t]
\centering
\includegraphics[width=0.7\textwidth,keepaspectratio]{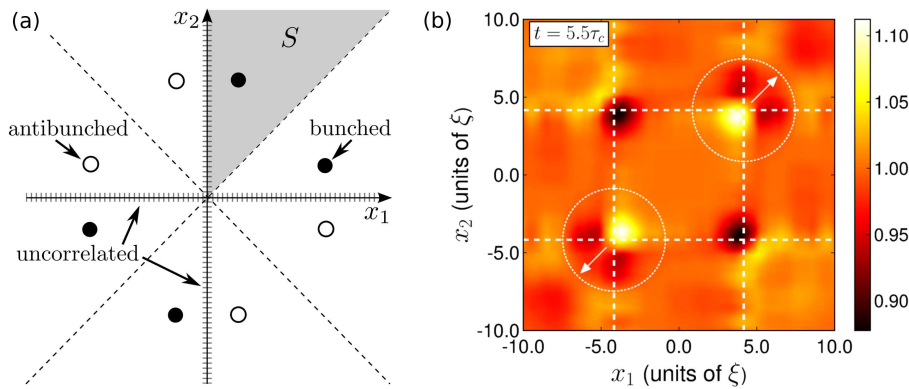}
\caption{(Color online)
(a) Structure of two-body correlations for the case $g_2(0,0)=1$. 
(b) $g_2(x_1,x_2)$ at $t=5.5\tau_c$ for a density-engineered
initial state of $N=100$ bosons
with $\gamma=0.04$ in a one-dimensional box of length $L=20\xi$
(obtained by (ML-)MCTDHB simulations).
The dashed lines indicate the positions of two counter-propagating 
gray solitons and the arrows point into their directions of motion. The circles
indicate the characteristic correlation pattern observed for a single gray
soliton in \cite{kronke_many-body_2014}. Strength of 
two-body correlations at the symmetry
center: $g_2(0,0)\approx1.003$.
}
\label{fig:g2_structure}
\end{figure*}
%%%%%%%%%%%%%%%%%%%%%%%%%%%%%%%%%%%%%%%%%%%%%%%%%%%%%%%%%%%%%%%%%%%%%%%%%%%%%%%%
\subsection{Reconstruction of even NO}\label{sec:rectr_even}
% {\em Reconstruction of even NO.---}
Due to the presence of the unknown $\Delta$ in (\ref{eq:rho1}), density
and density-fluctuation
measurements can only be employed to
relate the NO density difference $|\phi_0(x)|^2-|\phi_1(x)|^2$ at different 
points in space, $x=x_{1/2}$, but not to extract $|\phi_0(x)|^2$ itself. 
Nevertheless, using (\ref{eq:phi1_reconstr})
the possible candidates for
$|\phi_0(x)|^2$ can be restricted to a one-parametric family\footnote{In the
$g_2(0,0)\rightarrow1$ limit, this equation becomes equivalent to
(\ref{eq:rho1}) such that no information is gained.},
\begin{equation}\label{eq:phi0_reconstr}
 \frac{|\phi_0(x)|^2}{\rho_1(x)}=\left(\frac{1}{\Delta}+\big(1-\frac{1}{\Delta}\big)
\frac{g_2(0,0)-g_2(x,0)}{g_2(0,0)-1}\right),
\end{equation}
where $\Delta\in(0,1)$. Thus, a theoretical estimate for $\Delta$ by means of
e.g.\ number-conserving Bogoliubov theory
\mbox{\cite{girardeau_theory_1959,gardiner_particle-number-conserving_1997,
girardeau_comment_1998,
castin_low-temperature_1998,
bogoliubov_theory_of_bec_in_particle_representation_Sacha_Dziarmaga_PRA2003}} 
would allow for uniquely
determining the density of the even NO. 
In order to obtain a measurement protocol for $\Delta$
as an alternative, 
we inspect the first-order coherences
$g_1(x,-x)\equiv\rho_1(x,-x)/\sqrt{\rho_1(x)\rho_1(-x)}$ 
\cite{glauber_quantum_1963,naraschewski_spatial_1999}. Substituting $|\phi_{0/1}(x)|^2$
by (\ref{eq:phi0_reconstr}) and (\ref{eq:phi1_reconstr}), respectively,
in $\rho_1(x,-x)$,
we arrive at
\begin{equation}\label{eq:g1_g2_relation}
g_1(x,-x)=1-2(1-\Delta)\frac{g_2(0,0)-g_2(x,0)}{g_2(0,0)-1}.
\end{equation}
Thus, the additional knowledge of $g_1(x^*,-x^*)$ for some convenient
position $x^*$ 
with $g_2(x^*,0)\neq g_2(0,0)$ is
sufficient for determining $\Delta$.
In the case that we have only experimental access to the modulus of
$g_1(x,-x)$ but not to its
sign, we may extract $\Delta=\Delta_\pm(x)$ for
both signs, i.e.\ $\pm|g_1(x,-x)|$, from (\ref{eq:g1_g2_relation}) for all $x$
of finite density with 
$g_2(x,0)\neq g_2(0,0)$.
One easily
verifies $\Delta_+(x)\geq \Delta_-(x)$ and, in many situations,
the local sign of $g_1(x,-x)$ can then be fixed
by requiring $\Delta$ not to depend on $x$.
% % , since $\Delta$
% % must not depend on $x$, one may apply the following
% % criterion:
% % Given an interval where only one of the quantities $\Delta_\pm(x)$
% % stays constant, this locally constant quantity 
% % uniquely fixes the sign of $g_1(x,-x)$ on that interval and, thereby, 
% % determines $\Delta$.
Knowing $\Delta$ and an estimate for $N$, we may also infer 
${\rm var}(n_0)$ from (\ref{eq:g2_00}). Finally, Eq. (\ref{eq:g1_g2_relation})
 gives the conceptual insight
that the average
fraction of bosons in the even NO mediates a relationship between
the first order coherences $g_1(x,-x)$ and two-body correlations
$g_2(x,0)$.
%%%%%%%%%%%%%%%%%%%%%%%%%%%%%%%%%%%%%%%%%%%%%%%%%%%%%%%%%%%%%%%%%%%%%%%%%%%%%%%%
\subsection{Spatial structure of two-body correlations}\label{sec:struct_corr}
% {\em Spatial structure of two-body correlations.---}
While the two-body correlation function features a particle exchange
and a two-body parity symmetry,
it does not remain invariant under 
a parity operation acting on one atom only. By inspecting 
$\rho_2(x_1,x_2)+\rho_2(x_1,-x_2)-2\rho_1(x_1)\rho_1(x_2)$, i.e.\
essentially the sum of density-density correlations at $(x_1,\pm x_2)$,
the parities of the NOs can be 
employed to eliminate the off-diagonal term $\propto\alpha$ such that
a relationship between $g_2(x_1,x_2)$ and $g_2(x_1,-x_2)$ can be established.
Here, we have to distinguish two cases:
(i) In the absence of two-body correlations at the symmetry center, 
i.e.\ $g_2(0,0)=1$, we obtain the relation $g_2(x_1,x_2)+g_2(x_1,-x_2)=2$, which
has
three important consequences. Firstly,
the $g_2$ function is fully determined by its values in the sector
$S=\{(x_1,x_2)|{0\leq x_1\leq x_2}\}$ (cf.\ Fig.\ \ref{fig:g2_structure}(a) for
an
illustration).
Secondly, local bunching (anti-bunching) structures $g_2(x_1,x_2)>1$ 
($g_2(x_1,x_2)<1$) for $x_1\approx x_2$ translate into non-local anti-bunching
(bunching) structures of the same magnitude at $(x_1,-x_2)$. Thirdly, pairs
of atoms are uncorrelated on the $x_{1/2}$ axis, i.e.\ $g_2(x,0)=g_2(0,x)=1$.
(ii) In the presence of two-body correlations at the symmetry center, 
we may employ the reconstruction formula 
(\ref{eq:phi1_reconstr}) to
obtain the 
following functional equation, which has to be fulfilled for every
$g_2$ with $g_2(0,0)\neq1$ in order to be compatible with
the many-body wave function (\ref{eq:wfn}),
\begin{eqnarray}\label{eq:g2_func_eq}
g_2(x_1,x_2)+g_2(x_1,-x_2)=
2\Big(1+\frac{f(x_1)f(x_2)}{[g_2(0,0)-1]}\Big),
\end{eqnarray}
with $f(x)\equiv g_2(x,0)-1$.
This restriction on the functional form of $g_2$ may be used experimentally
as a necessary condition for testing the validity of the two-orbital
approximation.
%%%%%%%%%%%%%%%%%%%%%%%%%%%%%%%%%%%%%%%%%%%%%%%%%%%%%%%%%%%%%%%%%%%%%%%%%%%%%%%%
\subsection{Applications}\label{sec:appl}
% {\em Applications.---}
Dark solitons, being 
well-known for their stability within the mean-field approximation 
(see \cite{dark_solitons_in_atomic_BEC_from_theo_to_exp_Frantzeskakis_JPA2010}
and references therein), suffer from a quantum-fluctuations induced
decay due to an incoherent scattering of atoms from the soliton orbital 
into an orbital localized at the soliton position, c.f.\ e.g.\
\cite{quantum_depletion_of_dark_soliton_anomalous_mode_PRA2002,
quantum_entangled_dark_solitons_in_optical_lattices_Carr_PRL09,
delande_many-body_2014,kronke_many-body_2014}. Due to the increasing
population of this orbital, the depth of the characteristic 
minimum in the reduced one-body density is reduced, i.e.\ the soliton contrast
decreases on average over many absorption image measurements.
Since
this decay process can qualitatively be understood within a two-orbital 
approximation, dark solitons constitute a straightforward example for
testing the validity of the above insights in situations when further orbitals
participate with, however, minor weight. In the following, we consider
$N$ bosons of mass $m$ in a one-dimensional box potential of length $L$ with a
contact interaction strength $g$. This system is governed by the
Hamiltonian
$\hat H =\sum_i\hat p_i^2/2+\sqrt{\gamma}\sum_{i<j}\delta(\hat x_i-\hat x_j)$
in a unit system based on the chemical potential $\mu_0=gN/L$, the healing
length $\xi=\hbar/\sqrt{m\mu_0}$ and the
correlation time $\tau_c=\hbar/\mu_0$, where $\gamma=mgL/(\hbar^2N)$ denotes
the Lieb-Liniger parameter.

\begin{figure}[t]
\centering
\includegraphics[width=1.0\linewidth,keepaspectratio]{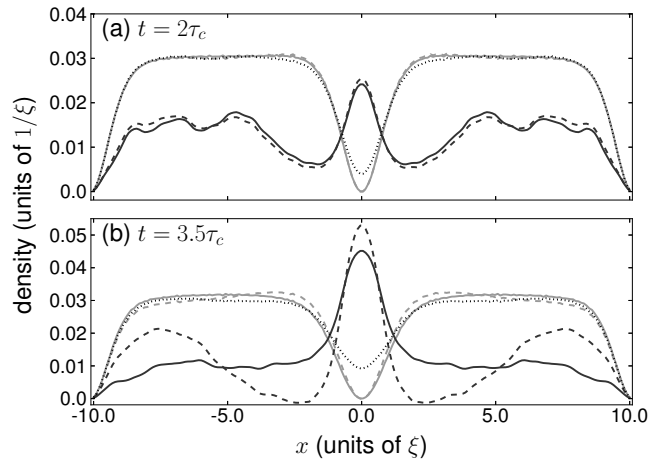}
\caption{
One-body density and density of the two most dominant NOs at times $t=2\tau_c$
(a) and $t=3.5\tau_c$ (b) for a many-body system initially featuring a
black soliton centered at $x=0$. All parameters are as in Fig.\
\ref{fig:g2_structure}(b). 
Dotted line: $\rho_1(x)$. Solid lines: Exact NO densities $|\phi_0(x)|^2$
(black) and $|\phi_1(x)|^2$ (gray, reduced by factor $2$). All these
curves are obtained from (ML-)MCTDHB {\it ab-initio} simulations. Dashed lines:
Corresponding reconstructions
of the NO densities. Estimated $\Delta\approx0.079$, $0.088$ for (a), (b),
respectively. Exact values: $\Delta\approx0.073$, $0.098$. 
Two-body correlations at the symmetry center:
$g_2(0,0)\approx2.699$, $2.013$.
}
\label{fig:norb_reconstr}
\end{figure}

Both for finding the initial state and for the subsequent propagation in
the following two scenarios,
we employ our recently developed Multi-Layer Multi-Configuration
Time-Dependent Hartree Method for Bosons \mbox{(ML-MCTDHB)}
\cite{kronke_non-equilibrium_2013,cao_multi-layer_2013}, which reduces to the
pioneering MCTDHB method \cite{MCTDHB_Cederbaum} if
applied to a single species in one spatial
dimension. This method is based on an expansion of the total many-body
wave-function with respect to bosonic number states with an underlying
time-dependent, dynamically optimized single particle basis of $M$ states. All
conceivable number state configurations for the given $M$ single-particle states
are into account. Being based on a variational principle, the (ML-)MCTDHB
equations provide us with a variationally optimized solution to the
time-dependent many-body Schr\"odinger equation. By incrementing $M$, we found
in both scenarios discussed below
that convergence is ensured on the considered time-scale if $M=4$
(cf.\
\cite{kronke_many-body_2014}). A NO analysis of the
full numerical data reveals that two NOs contribute with significant weight
while the probability to find an atom in one of the other two NOs does not 
exceed $0.023$ for the time-scales of Fig.\ \ref{fig:g2_structure}(b)
and Fig.\ \ref{fig:norb_reconstr}.

First, we start with the ground state of $N=100$ atoms in the box
with an additional
Gaussian barrier $V(x)=h\exp[-x^2/(2w^2)]$, $h=60\mu_0$,
$w\approx0.07\xi$ so that we engineer a pronounced density notch at
$x=0$. Having switched off the barrier, we let the interacting
many-body system evolve in the box potential. In the course 
of time, the single density minimum splits into a pair of counter-propagating
gray solitons, which are slowly decaying due to quantum-fluctuations
\cite{kronke_many-body_2014}.
We have shown that a single gray soliton
is
accompanied by highly localized
two-body correlations in the vicinity of the instantaneous soliton position
resulting in a bunching of atoms in the soliton flank opposite to its direction
of movement (cf.\ circles in Fig.\
\ref{fig:g2_structure}(b)). Yet the $g_2$
function of two counter-propagating gray solitons turns out to be more than 
the sum of the local correlation patterns of the individual gray solitons - 
additional non-local correlations occur between the two solitons
(Fig.\ \ref{fig:g2_structure}(b) and \cite{kronke_many-body_2014}).
Observing numerically that $|g_2(0,0)-1|\ll1$, we may now understand these 
non-local correlations as a generic property of parity-symmetric systems 
with essentially two occupied NOs: Denoting the position of the soliton
moving to the right/left with $x_{R/L}$, our above results imply
$g_2(x_R-\epsilon,x_L+\epsilon)\approx 2 -g_2(x_R-\epsilon,x_R-\epsilon)$
such that bunching in the back of a single soliton (${\epsilon>0}$)
translates into antibunching of approximately same magnitude 
for finding an atom each in the back of each soliton.

Second, in order to realize a situation with significant deviations of
$g_2(0,0)$ from
unity, we additionally imprint a relative phase of $\pi$ between the two
half-spaces at $t=0$ \cite{
quantum_depletion_of_dark_soliton_anomalous_mode_PRA2002,
quantum_entangled_dark_solitons_in_optical_lattices_Carr_PRL09,
delande_many-body_2014,kronke_many-body_2014}. Thereby, a black soliton
is initialized at $x=0$.  As time evolves, the density minimum becomes
filled up by incoherently scattered atoms inducing strong bunching
correlations at the symmetry center. Fig.\ \ref{fig:norb_reconstr} clearly shows
that 
the density of the dominant NO of odd parity, $|\phi_1(x)|^2$, features
the characteristic density notch of a black soliton and can be reliably 
reconstructed by the scheme (\ref{eq:phi1_reconstr}) at times when
the soliton contrast in the full density $\rho_1(x)$ has been reduced. For
longer times, the reconstructed $|\phi_1(x)|^2$ deviates slightly more from
the full numerical results
since the third and fourth dominant NO have gained more population. The
reconstruction of $|\phi_0(x)|^2$, i.e.\ the NO mostly responsible for the
soliton decay and strong two-body correlations, turns out to be more sensitive
to the slight population of these NO:
Due to the phase-imprinting scheme, we expect
most of the atoms to reside in an odd NO.
Finding numerically $\Delta_{-}(x)<1/2<\Delta_{+}(x)$, we thus take the negative
sign of $g_1(x,-x)$ and estimate $\Delta$ by averaging $\Delta_{-}(x)$ 
% (which
% depends on $x$ because of the slight population of the third and fourth NO) 
over 
some interval. As a result, we can fairly well reconstruct $|\phi_0(x)|^2$
according to (\ref{eq:phi0_reconstr}) for not too long times (Fig.\
\ref{fig:norb_reconstr}(a)). Thus, our scheme can be used to
experimentally verify the microscopic decay mechanism of a black soliton
contrast in the reduced one-body density
via a NO being localized at the position of the soliton, given that
thermal excitations are sufficiently suppressed as achievable in 
nowadays experiments \cite{weld_spin_2009}. At longer times ($t\gtrsim
3\tau_c$, see Fig.\
\ref{fig:norb_reconstr}(b)), i.e.\ when the
assumption of only two occupied orbitals becomes less valid, however, the
reconstructed $|\phi_0(x)|^2$ deviates much stronger from the full
numerical results compared to the reconstruction of $|\phi_1(x)|^2$.

The reconstruction formulas (\ref{eq:phi1_reconstr}) and
(\ref{eq:phi0_reconstr}) have been derived under the assumption of perfect
experimental resolution. In order to test the robustness of the
reconstruction recipe against
finite resolution, we have (i) embedded our numerical simulation results 
for the considered purely one-dimensional model into three-dimensional
coordinate space by
assuming all atoms to reside in the ground state
of the transverse harmonic oscillator and (ii) convoluted the one- and
two-body densities with a Gaussian point-spread-function of variable width
$\sigma$. These coarse-grained quantities have then been inserted in the
reconstruction
formulas (\ref{eq:phi1_reconstr}) and
(\ref{eq:phi0_reconstr}). We have found that the NO
densities
and thereby the decay mechanism of the soliton contrast in the reduced
one-body density can be reconstructed as long as $\sigma \lesssim\xi/4$,
which was to be expected (plots not shown). This requirement on the optical
resolution is demanding, of course, but not out of reach since the healing
length could be raised above the resolution limit by a Feshbach resonance, for
example. For such values of $\sigma$, the reconstructed $|\phi_0(x)|^2$ density
happens to be sufficiently robust against errors in the depletion $\Delta$, for
which we
assumed a relative uncertainty of $20\%$. Finally, we remark that not
$\rho_2(x_1,x_2)$ but density-density correlations $
\langle [\hat \psi^\dagger(x_1)\hat \psi(x_1)-N\rho_1(x_1)]
[\hat \psi^\dagger(x_2)\hat \psi(x_2)-N\rho_1(x_2)]\rangle$ are measured in
experiments. Due to the canonical commutation relations for bosonic field
operators, however, these two quantities essentially differ by an
autocorrelation peak $\propto\delta(x_1-x_2)$, which becomes smooth when
convoluted with a point-spread-function and whose contribution is suppressed as
$1/N$ for large particle numbers.

%%%%%%%%%%%%%%%%%%%%%%%%%%%%%%%%%%%%%%%%%%%%%%%%%%%%%%%%%%%%%%%%%%%%%%%%%%%%%%%%
\section{Conclusions}\label{sec:concl}
% {\em Conclusions.---}
We have shown how physical knowledge about the structure
of the many-body wave-function can be employed for deriving generic 
properties of two-body correlations and an experimentally relevant 
reconstruction scheme for the NO densities. In addition to
the discussed dark soliton example, our results should 
be applicable to many other systems such as Bose gases in a double-well potential
\cite{a_bosonic_josephson_junction_Oberthaler_JPB2007}, fragmenting bright
solitons
\cite{soliton_fragmenton,swift_loss_coherence_Cederbaum_PRL2011} or
symmetrically 
colliding fragments in a harmonic trap
\cite{streltsova_interferometry_2014,sakmann_single_2015}. If
the central density turns out to be vanishing or
too small such that $g_2(0,0)$ becomes effectively ill-defined, 
the whole analysis has to be carried out in momentum space via 
long ToF measurements. 
We hope that our work stimulates the interest in NO reconstruction schemes
such that these microscopic quantities become experimentally accessible
for a broad class of systems.
%%%%%%%%%%%%%%%%%%%%%%%%%%%%%%%%%%%%%%%%%%%%%%%%%%%%%%%%%%%%%%%%%%%%%%%%%%%%%%%%
\begin{acknowledgments}
We would like to thank T. Schumm, H. Moritz, J. Simonet and 
L. Cao for fruitful discussions 
as well as J. Kn\"orzer and J. Stockhofe for helpful comments on the manuscript.
S.K. gratefully acknowledges a scholarship by the Stu\-dien\-stif\-tung des
deutschen Volkes. P.S. gratefully acknowledges financial support by the
Deutsche Forschungsgemeinschaft in the framework of the SFB 925 ``Light induced
dynamics and control of correlated quantum systems''.
\end{acknowledgments}
%%%%%%%%%%%%%%%%%%%%%%%%%%%%%%%%%%%%%%%%%%%%%%%%%%%%%%%%%%%%%%%%%%%%%%%%%%%%%%%%
\bibliography{no_tomography_lit_no_title}
% \bibliography{apssamp}
\bibliographystyle{unsrt}

\end{document}